\documentclass[aps,pre,twocolumn]{revtex4}
\usepackage{graphicx}
\begin{document}

\title{Model for erosion-deposition patterns}

\author{D. O. Maionchi$^{1,2}$, A. F. Morais$^{1}$, R.~N.~Costa Filho$^{1,3}$,
J.~S.~Andrade Jr.$^{1,4}$, H. J. Herrmann$^{1,4}$}
\affiliation{$^{1}$Departamento de F\'isica, Universidade Federal de
Cear\'a, 60451-970 Fortaleza-CE, Brazil} \affiliation{$^{2}$Institut
f\"{u}r Computerphysik, Universit\"{a}t Stuttgart, Pfaffenwaldring
27, 70569 Stuttgart, Germany} \affiliation{$^{3}$The Abdus Salam
International Centre for Theoretical Physics, Strada Costiera 11,
34014 Trieste, Italy} \affiliation{$^{4}$Computational Physics, IfB,
ETH H\"{o}nggerberg, HIF E 12, CH-8093 Z\"{u}rich, Switzerland}

\date{\today}

\begin{abstract}
We investigate through computational simulations with a pore network
model the formation of patterns caused by erosion-deposition
mechanisms. In this model, the geometry of the pore space changes
dynamically as a consequence of the coupling between the fluid flow
and the movement of particles due to local drag forces. Our results
for this irreversible process show that the model is capable to
reproduce typical natural patterns caused by well known erosion
processes. Moreover, we observe that, within a certain range of
porosity values, the grains form clusters that are tilted with respect
to the horizontal with a characteristic angle. We compare our results
to recent experiments for granular material in flowing water and show
that they present a satisfactory agreement.
\end{abstract}


\maketitle

\section{Introduction}

Nature preservation and environmental protection are important issues
on the agendas of governments and non-governmental
organizations. Consequently, these issues have been the subject of
intense study, where the main concern is to understand how the actions
of human beings affect the environment. For example, deforestation and
pollutant emissions are related to climate changes resulting in floods
and erosion. In particular, erosion can be responsible for diminishing
the quality of life, because it affects the soil causing a negative
impact on the economy. Aside from its economical and ecological
aspects, the erosion problem also attracts the interest of geologists
and physicists. In geology, this is an extremely rich area as many of
the patterns observed in nature stem from erosion or deposition
processes. In physics, the formation of such patterns span a huge
range of spatial and temporal scales. This pattern formation process
is directly related to the transport of solid granular particles via a
fluid and presents a rich phenomenology along with a variety of
applications \cite{Robert:1999,Jaeger:1996}. Particular applications
are fractal river basins, meandering rivers, dune fields, granular
avalanches, and ripple marks on sand banks or on coastal continental
platforms.

\begin{figure}[t]
\includegraphics[width=70 mm]{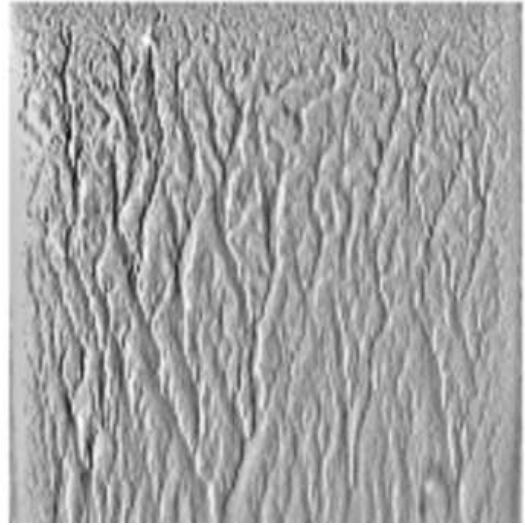}
\caption{Pattern observed in erosion an experiment with commercial 
abrasive powder. Angles between $30^{\circ}$ and $90^{\circ}$ were 
found for different chevron alignments \cite{Adrian:2003}.}
\label{experim}
\end{figure}

It is notoriously difficult to provide a fully consistent description
of particle laden flows either from a one-phase or a two-phase point
of view. Most of the practical knowledge of erosion comes from
empirical laws often derived from field measurements.  This has
provoked interest in theoretical descriptions of these systems
\cite{Herrmann2:2002,Herrmann1:1999,Ertas:2002}
and  visualization of them in computational simulations. Several
attempts to understand the dynamics of river basin formation from
the statistical physics point of view have been made recently, but
many questions are raised when one tries to relate basic transport
properties to large-scale pattern forming instabilities
\cite{Seybold:2007}. A fundamental open question is the following: how
do objects made of granular materials respond to the action of
external factors? Many experiments
\cite{Adrian:2003,Pouliquen3:1999,Igor:2006,Pouliquen:1999,Tamas:2005,Pouliquen4:2003,Pouliquen1:2004,Midi:2004}
have been performed in the last few years to answer this question.
For example, in Fig.~\ref{experim}, we see a laboratory-scale
experiment which reproduces a rich variety of natural patterns with
few control parameters \cite{Adrian:2003}. These patterns are
characterized by chevron alignments, what means that the paths
created by the fluid present a characteristic angle that depends on
the parameters varied in the experiment.

In this paper, we investigate through numerical simulation a physical
model that is designed to represent the generic situation of flowing
water on a plane composed of erodible sediment layer. This occurs
naturally when the sea retreats from the shore or when a reservoir is
drained. The flow of granular materials on planes is also of interest
within the context of both industrial processing of powders and
geophysical instabilities such as landslides and avalanches. These
flows have been found to be complex, exhibiting several different flow
regimes as well as particle segregation effects and instabilities
\cite{Savage:1997}. This paper is organized as follows. In Section II 
we describe the model used to simulate the erosion-deposition
patterns. These patterns are presented and analysed in Section III,
while the conclusions are left for Section IV.

\begin{figure}[t]
\includegraphics[width=70 mm]{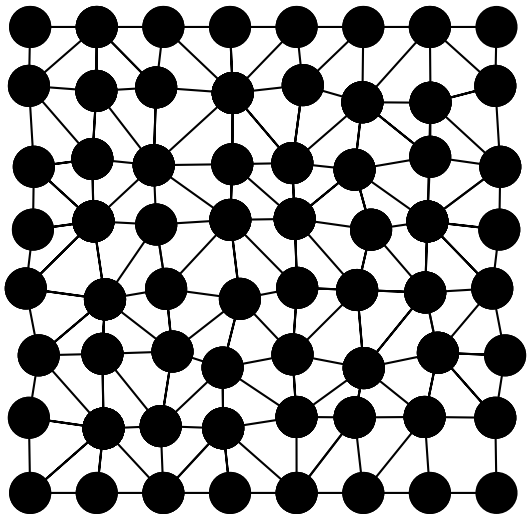}
\caption{A typical initial configuration of the regularized random 
network with $64$ grains and its corresponding triangulation.}
\label{confinicial}
\end{figure}
\begin{figure}[b]
\includegraphics[width=80 mm]{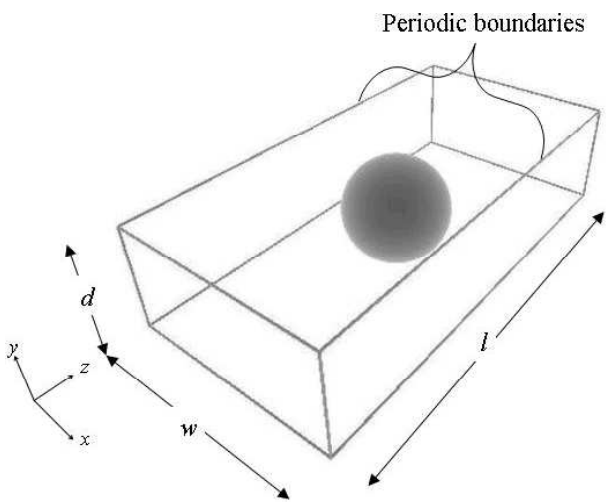}
\caption{The velocity field is determined in the channel of length
$l$ and width $w$ with one sphere of diameter $d$ inside. The
mean velocity of the fluid $v_f$ is calculated in the rectangular
cross-section.}\label{ara}
\end{figure}

\section{Model Formulation}

We model a system that takes into account the interaction of a
granular medium with an incompressible Newtonian fluid flowing through
the corresponding pore space. The granular medium is initially
considered as a $N \times N$ regularized random network (RRN) on a
plane where the sites are the centers of mass of spherical grains with
diameter $d$ that are totally submerged in a fluid (e.g., water). We
initialize this network as a regular square where the distance between
the closest neighbors is $l_{0}$. The points (centers of mass of the
grains) are then moved randomly along vectors with arbitrary direction
and random magnitude that are smaller than the distance between the
points. In this way, the points are distributed randomly, but with a
characteristic distance. More precisely, in order to avoid the
occurrence of overlapping grains, the maximum value adopted for the
modulus of these dislocation vectors is $(l_{0}-d)/2$. Although the
lattice construction is made in $2D$, we are actually describing one
layer of a three-dimensional system, since the centers of mass of the
particles still lay on a $2D$ plane but the grains are considered to
be spheres. This association will enable us to generate a network of
capillaries representing a complex geometry of the pore space. At this
point, the entire system is triangulated considering each grain as a
vertex of a Voronoi tesselation. In Fig.~\ref{confinicial} we show a
typical initial RRN configuration and its corresponding triangulation

Next, we assume that the local pore geometry between each two nearest
neighbors grains $i$ and $j$ in this lattice can be modeled as a
capillary channel of length $l$ (the distance between barycenters of
the corresponding adjacent triangles), height $d$ and width $w$ equal
to the distance between their centers. As shown in Fig.~\ref{ara}, if
we consider periodic boundary conditions (PBC) in the $x$-direction,
such a channel should be equivalent to a parallelepiped of fluid
containing in its center a solid sphere of diameter $d$ (grain).
\begin{figure}[t]
\includegraphics[width=70 mm]{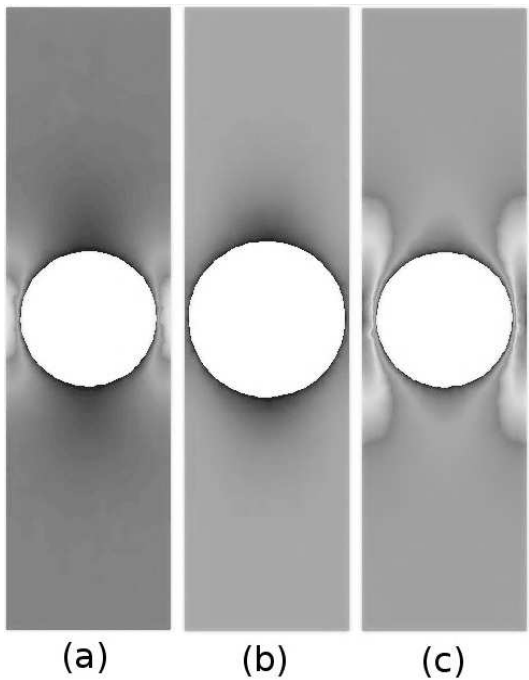}
\caption{Contour plots of the velocity field obtained numerically for 
$w/d=1.05$ at three different transverse planes of the channel,
namely, (a) $y/d=1/4$, (b) $y/d=1/2$ and (c) $y/d=3/4$. The gray
shades ranging from dark to light correspond to low and high velocity
magnitudes, respectively.} 
\label{color}
\end{figure}
\begin{figure}[t]
\includegraphics[width=70mm,angle=-90]{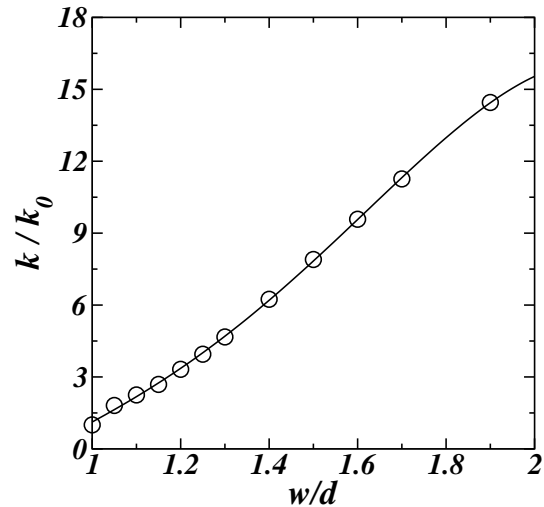}
\caption{Permeability $\kappa$ versus the ratio $w/d$ for the
local channel configuration shown in Fig.~(\ref{ara}). Here $\kappa_0$
is the permeability for a channel with ratio $w/d=1$. The solid line
is the best fit to the data of the fourth degree polynomial used to
interpolate the local permeability of the channels in the pore network
model.}
\label{k2}
\end{figure}

Here the Navier-Stokes and continuity equations in the
three-dimensional channel are solved using the commercial CFD software
FLUENT \cite{fluent}. We consider no-slip boundary conditions at the
bottom and assume that the top surface is shear-stress free. A
pressure gradient $\Delta p $ is imposed between the two ends 
of the channel in the $z$-direction. Its magnitude is sufficiently
small to ensure viscous flow conditions, i.e., a low Reynolds number
regime of flow. We adopt a non-structured tetrahedral mesh to
discretize the channel and an upwind finite-difference scheme is set
to perform the numerical simulations. The steady-state velocity and
pressure fields are calculated for different channel geometries by
systematically varying the ratio $w/d$ in the range $1.05 < w/d <
1.9$. 

In Figs.~\ref{color}a-c we show three contour plots of the velocity
field computed for a channel with porosity $w/d=1.05$ at heights
$y/d=4$, $2$ and $3/4$, respectively. Considering the low Reynolds
number conditions used in the simulations, the flow in the channel can
be characterized in terms of a permeability index $\kappa$ through the
relation,
\begin{equation}
v_{f}=-\frac{\kappa}{\mu}\frac{\Delta p}{\Delta z}~,
\label{darcy}
\end{equation}
where $\mu$ is the viscosity of the fluid and $v_{f}$ is the average
flow velocity. In practical terms, once the velocity and pressure
fields are obtained for a channel with a given value of $w/d$, we can
compute the mean velocity value $v_{f}$ through a cross-section
orthogonal to the flow in the system. By repeating this procedure for
different values of the overall pressure drop $\Delta p$, we first
confirm the validity of the linear relationship $v_{f} \propto \Delta
p$ as expected from Eq.~(\ref{darcy}), so that the permeability $\kappa$
can be directly calculated from the slope of the corresponding straight
line. As shown in Fig.~\ref{k2}, the dependence of $\kappa$ on the
ratio $w/d$ can be fully described as $\kappa/\kappa_{0}=f(w/d)$ where
$f(w/d)$ is a fourth degree polynomial of $w/d$ and $\kappa_{0}$ is the
permeability for a channel with unitary aspect ratio, $w/d=1$.
 
Once the local geometry and permeability of all capillaries in the
system is determined, we proceed by applying a constant pressure drop
between the inlet and outlet, i.e., the top and bottom of the entire
pore network. Periodic boundary conditions are assumed in the lateral
direction of the network, and the following local mass conservation
equations are imposed at each of their $N$ nodes to allow water flow
throughout the entire pore space:
\begin{equation}
\sum_{j} g_{ij}(p_{i}-p_{j})=0~~\text{for}~~i=1,2,...,N~,
\label{eq:conservation}
\end{equation}
where the index $j$ runs over all the neighbor nodes of node $i$,
$g_{ij} \equiv wd \kappa/l$ is the hydraulic conductance of the pore
and $p_{i}$ and $p_{j}$ are the pressures at nodes $i$ and $j$,
respectively. Equation~(\ref{eq:conservation}) corresponds to a set of
$N^{2}$ coupled linear algebraic equations that are solved in terms of the
nodal pressure field by means of a standard subroutine for sparse
matrices. From the pressures in the nodes, the velocity magnitude of
the fluid in each capillary can be computed.

\begin{figure}[t]
\includegraphics[width=60 mm]{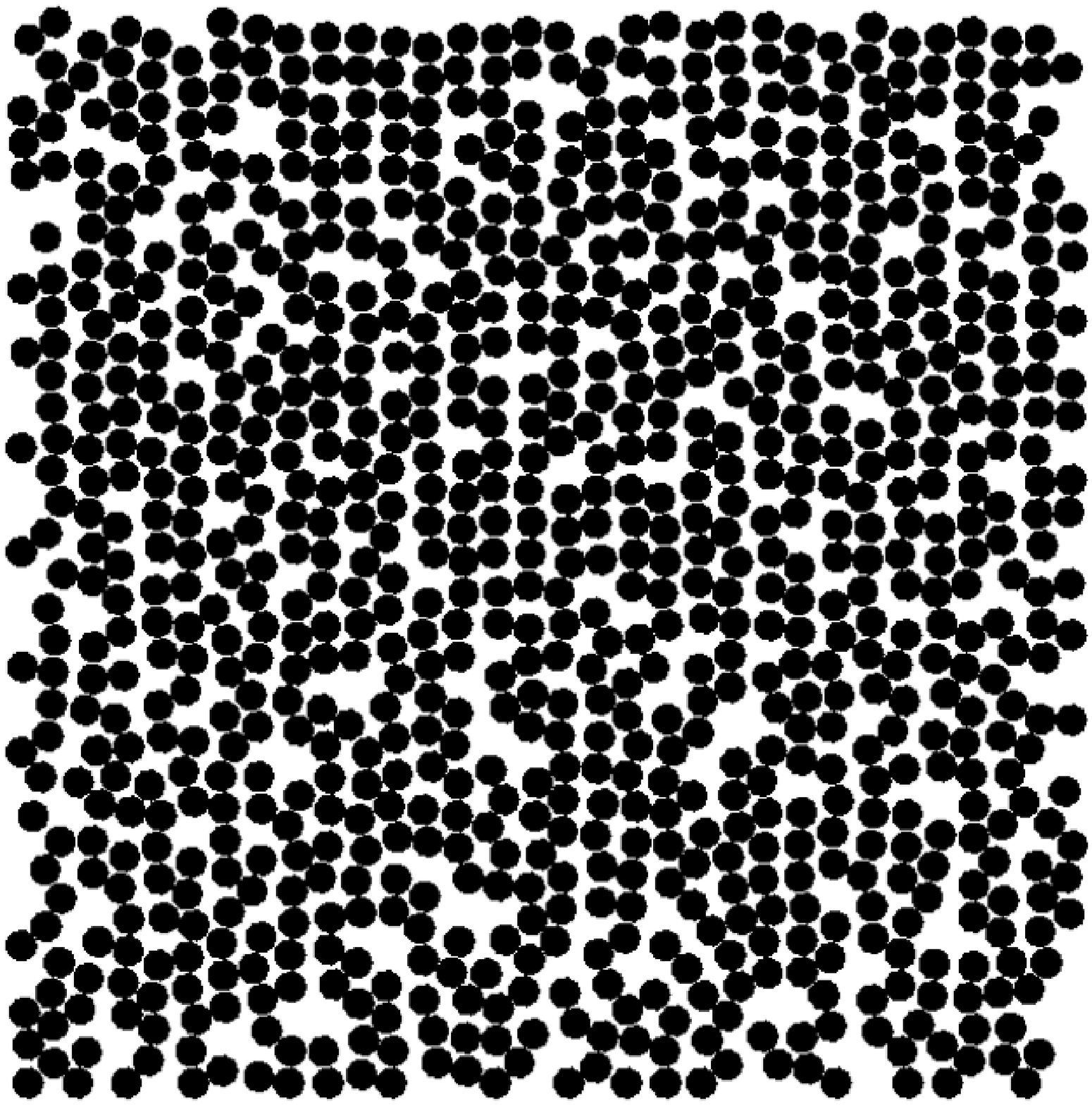}(a)
\includegraphics[width=60 mm]{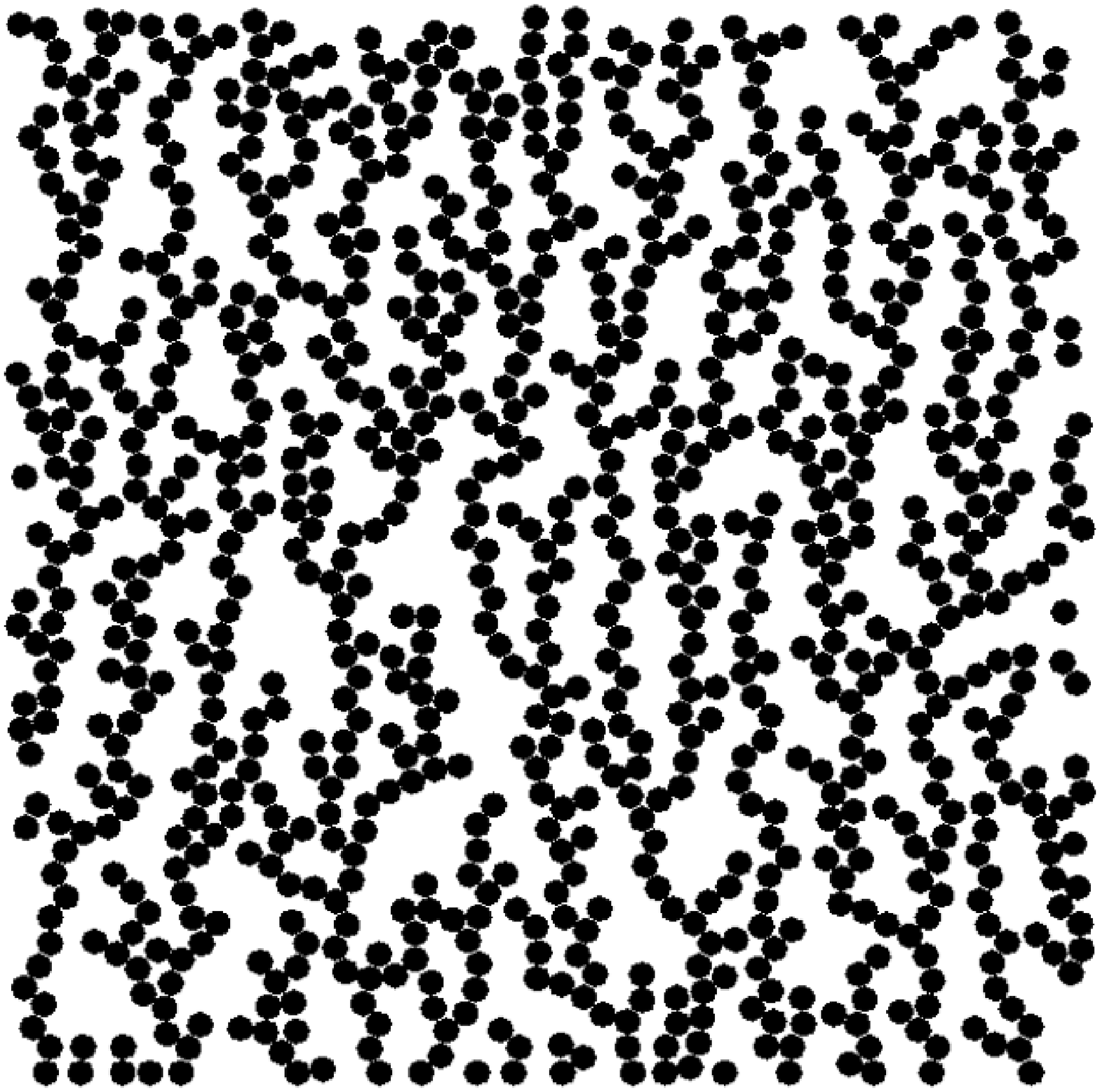}(b)
\includegraphics[width=60 mm]{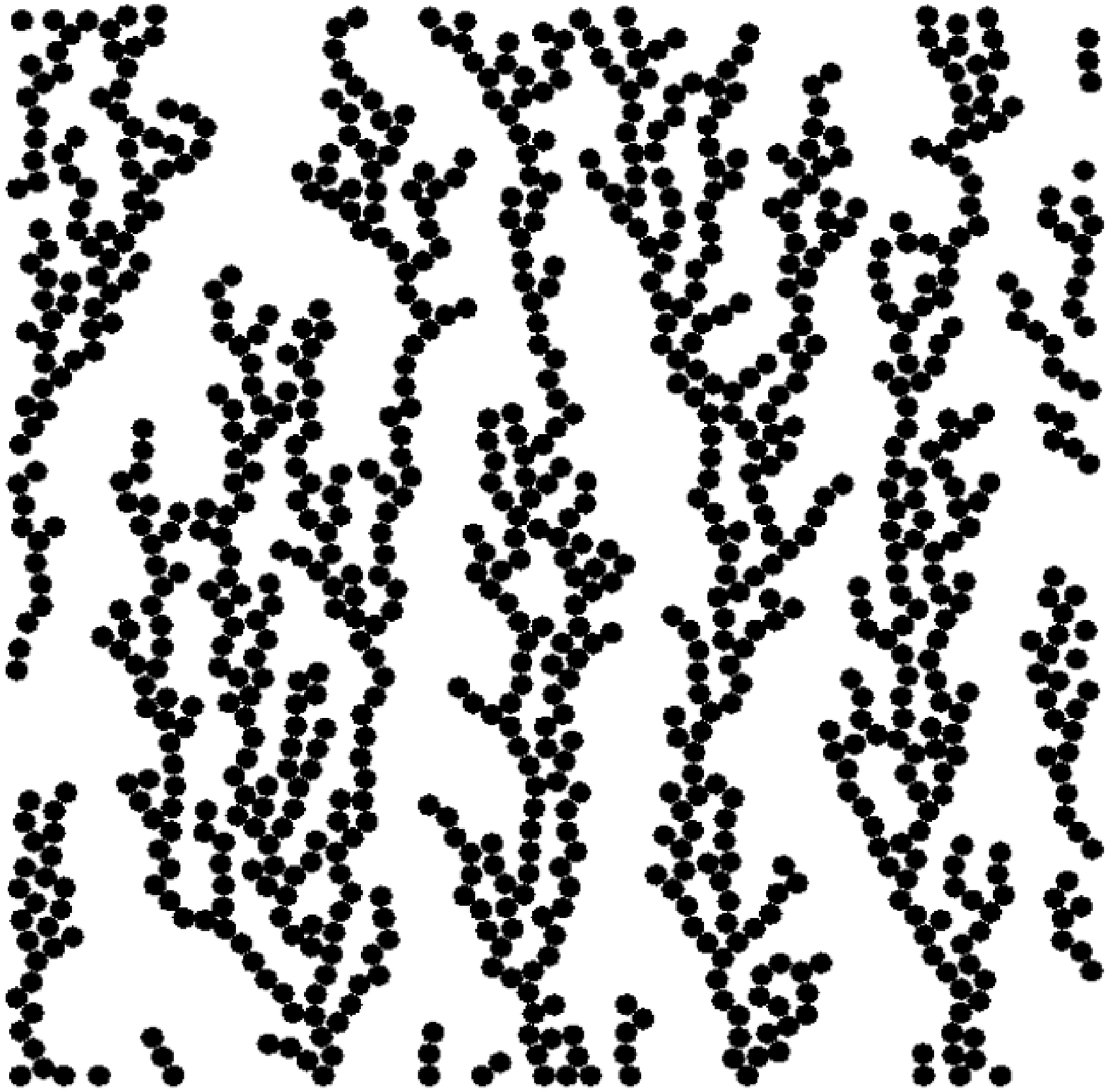}(c)
\caption{In (a), (b) and (c) we show the final configurations of the system
for porosities $\phi$=$0.57$, $0.71$ and $0.80$, respectively after
$11000$, $30000$ and $140000$ time steps. The grains arrange themselves
in positions that give rise to characteristic patterns.}
\label{conf}
\end{figure}

After the velocity field in the pore network is calculated, we allow
each grain to move in the system. By assuming that there is no
friction on the ground and drag is the only relevant force acting on
the particles we obtain, 
\begin{equation}
{m \frac{d^{2}\bf{x}}{dt^{2}}}=3 \pi \mu d \sum_{i} 
({\bf v}_{f}^{i}-{\bf v})~
\label{move}
\end{equation}
where ${\bf v}_{f}^{i}$ is the fluid velocity at the channel {\it i},
$\bf{v}$ is the particle velocity, and the vectorial sum on the right
is taken over all $n$ channels surrounding the particle. By
straightforward integration of the equation of motion (\ref{move}),
the velocity and displacement caused by the fluid drag to each grain
in the system during a time interval $\Delta t$ can be written as,
\begin{equation}
{\bf v}=\frac{\sum_{i}{\bf v}_{f}^{i}}{n}-\left(\frac{\sum_{i}{\bf v}_{f}^{i}}{n}
-{\bf v}_{0}\right)e^{-n C \Delta t}
\end{equation}
\begin{equation}
\Delta{\bf x}= \frac{\sum_{i}{\bf v}_{f}^{i}}{n}\Delta t+\left(\frac{\sum_{i}{\bf v}_{f}^{i}}{n}
-{\bf v}_{0}\right)\frac{e^{-n C \Delta t}}{n C}
\end{equation}
where $C=18 \mu/\rho d^{2}$, $\rho$ is the density of the grain,
and ${\bf v}_{0}$ is its velocity in the previous time step.

In our simulations, the center of each grain is displaced by
$\Delta{\bf{x}}$ using a time step $\Delta{t}$ that is sufficiently
small to numerically ensure that the pattern evolution remains
invariant when compared with results performed using even smaller time
steps. The distance between the top and bottom lines of the system is
kept constant with its center of mass moving according to the mean
displacement of all grains. What is crucial for the steadiness of the
pattern formation is that one grain cannot overlap with another
grain. After computing the movement of all grains at each time step,
the pore space is then modified, and we repeat the calculation of the
velocity field to move the grains again, and so on. The simulation
stops when the system reaches the steady-state, i.e., when the
geometry of the aggregate remains unchanged with time.

\section{Results}

We performed simulations on a lattice with $40 \times 40$ grains with
diameter $d=30 \mu m$ and density $\rho=2.75 g/cm^3$. These particles
are surrounded by water, i.e., a fluid of viscosity $\mu=10^{-3}Pa.s$
which flows at small Reynolds conditions ($Re \ll 1$) through a pore
space of porosity that can be varied in the range $0.6 < \phi
<0.8$. For each value of porosity we obtain results for five different
realizations of the initial random pore space.

In Figs.~\ref{conf}a-c we show three final stable configurations of
the model for different porosities values, $\phi$=$0.57$, $0.71$ and
$0.80$, respectively. As can be observed, the steady-state patterns
depend strongly on the porosity of the system. For sufficiently large
values of $\phi$, the occurrence of particle clusters in the form of
dendrites reflects the the strong coupling between fluid dynamics and
grain movement, where the aligned preferential channels for flow leads
to a high overall permeability of the porous system. For small
porosities, however, no characteristic pattern is observed. This is to
be expected since in compacted systems the grains do not have much
mobility while in loose systems the particles have the freedom to move
in almost all directions.

\begin{figure}[t]
\includegraphics[width=70mm,angle=-90]{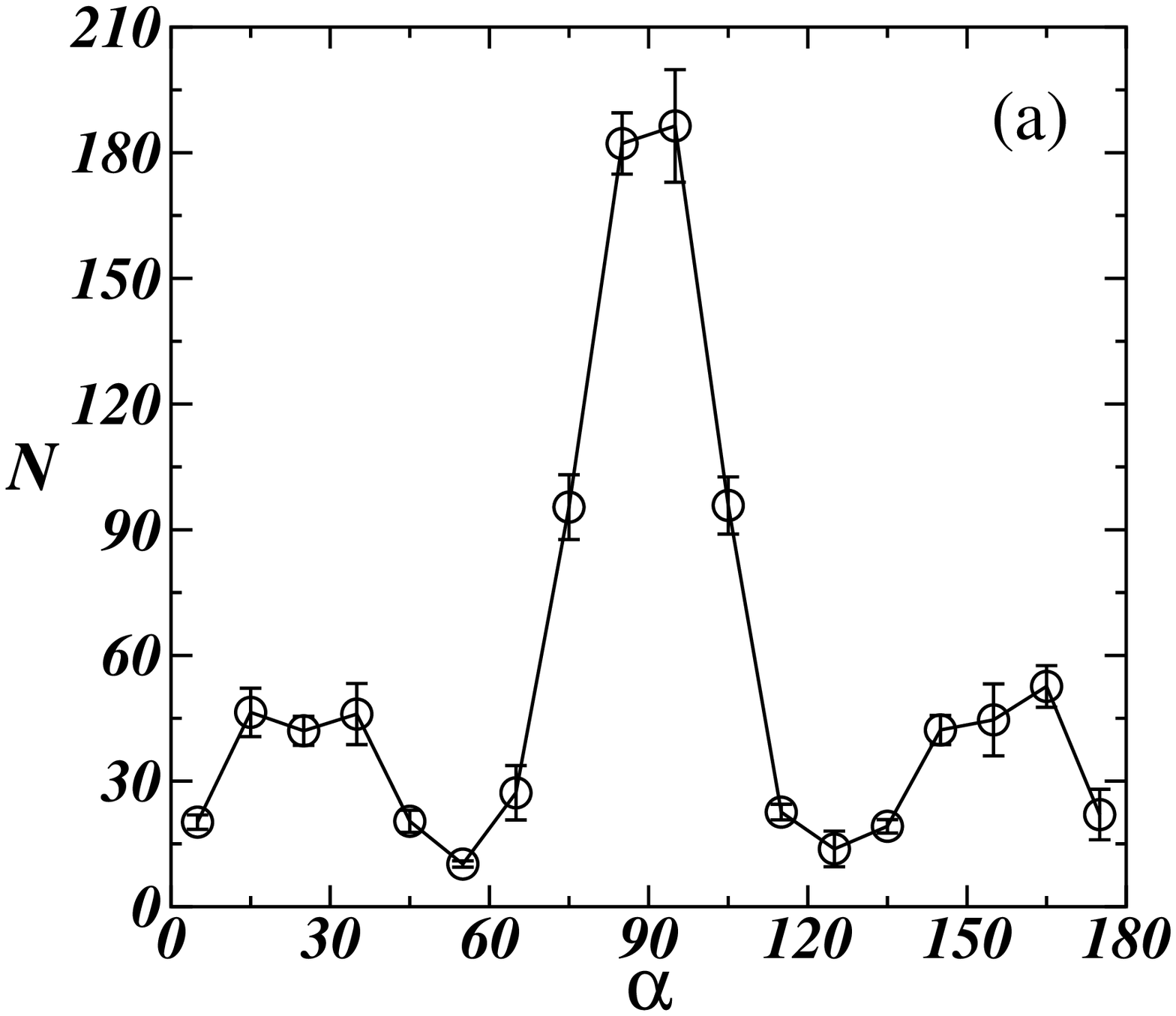}
\includegraphics[width=70mm,angle=-90]{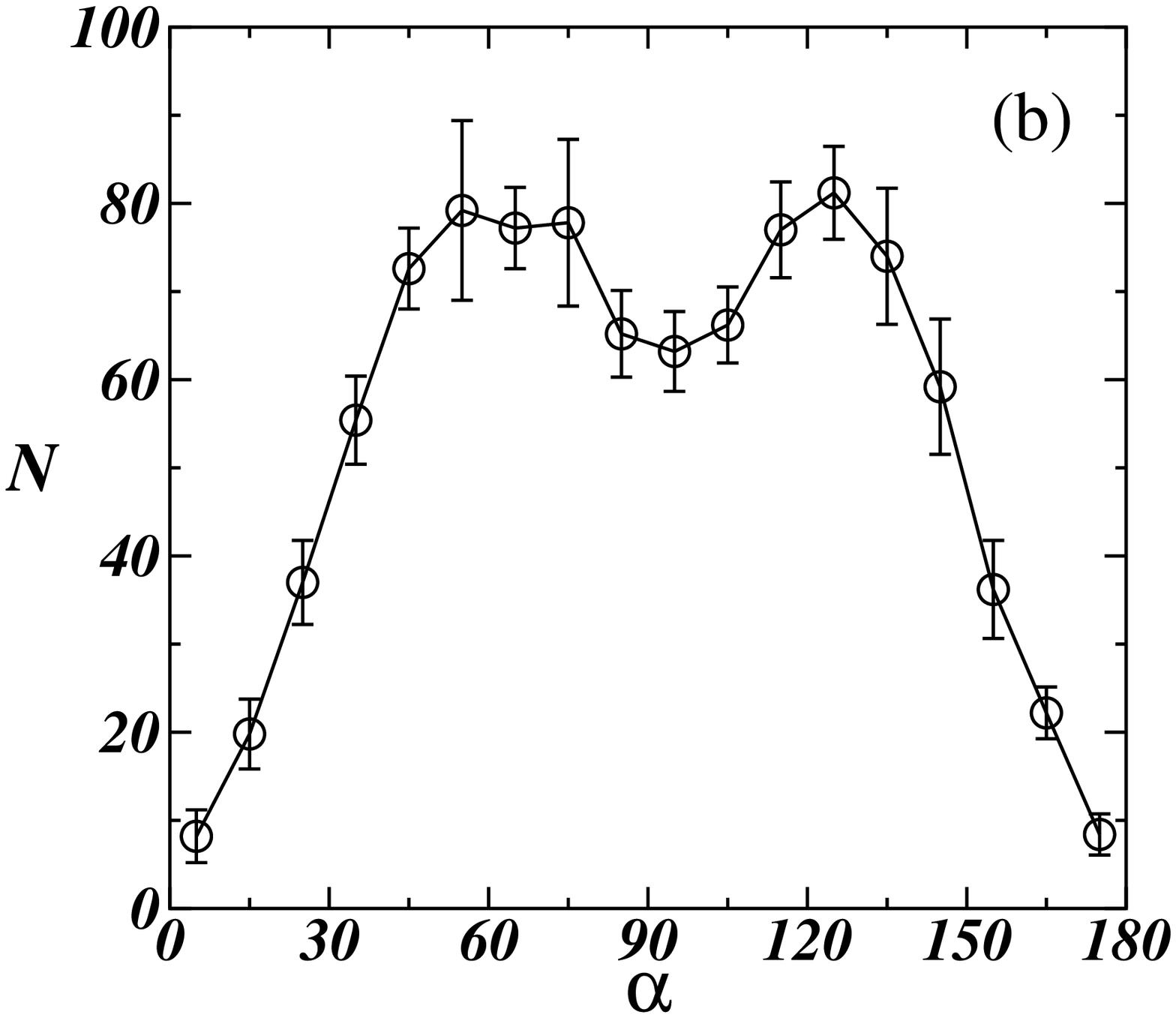}
\includegraphics[width=70mm,angle=-90]{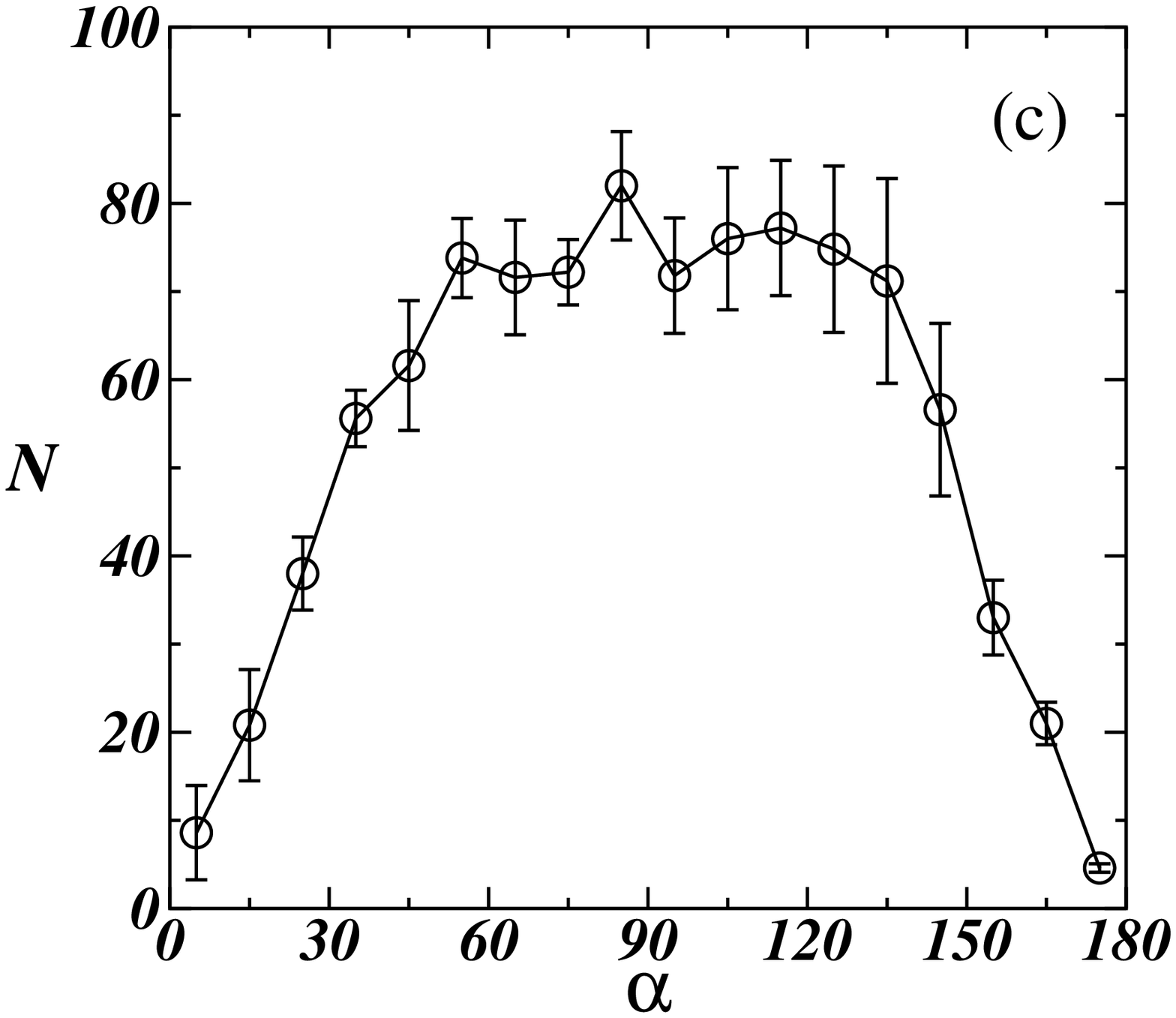}
\caption{Histograms of pairs of grains that are touching and for which
the line joining their centers of mass forms an angle $\alpha$ with
the axis $x$. These results represent the average over five
realizations of different porosities, namely, (a) $\phi=0.57$, (b)
$0.71$ and (c) $0.80$.}
\label{grud}
\end{figure}

The tendency to form a dendritic pattern in which the particles align
in preferential directions can be statistically quantified if we
determine for each pair of grains the angle $\alpha$ between the line
connecting their centers of mass and the direction orthogonal to the
flow (i.e., the $x$-direction shown in Fig.~\ref{conf}a). In
Figs.~\ref{grud}a-c we show the histograms of these angles $N(\alpha)$
for $\phi$=$0.57$, $0.71$ and $0.80$. For a low porosity system,
$\phi=0.57$, the results shown in Fig.~\ref{grud}a indicate that a
significant number of grains are aligned around $\alpha=25^{\circ}$
(and the symmetric direction of $155^{\circ}$), although the most
frequent angle lies in the vicinity of $90^{\circ}$. This means that
the particles tend to be aligned in the vertical $y$-direction (i.e.,
the direction of the flux), what is expected as the system cannot
change much from its initial configuration. In systems with
intermediate porosity values, $0.65 < \phi < 0.75$, there is a
substantial change in the preferred angle of alignment, which is
around $\alpha=60^{\circ}$ (and the symmetric direction of
$120^{\circ}$). This behaviour, as exemplified here in
Fig.~\ref{grud}b, resembles the chevron alignment reported in
Ref.~\cite{Adrian:2003}, where the results of the angle histograms are
presented for a porosity $\phi=0.71$. In Fig.~\ref{grud}c we show the
histogram $N(\alpha)$ for a large porosity system, $\phi = 0.80$,
where no evident preferential direction can be observed, with the
particles aligning themselves in angles between $50^{\circ}$ and
$130^{\circ}$, with approximately the same probability.

\begin{figure}[ht]
\includegraphics[width=70mm,angle=-90]{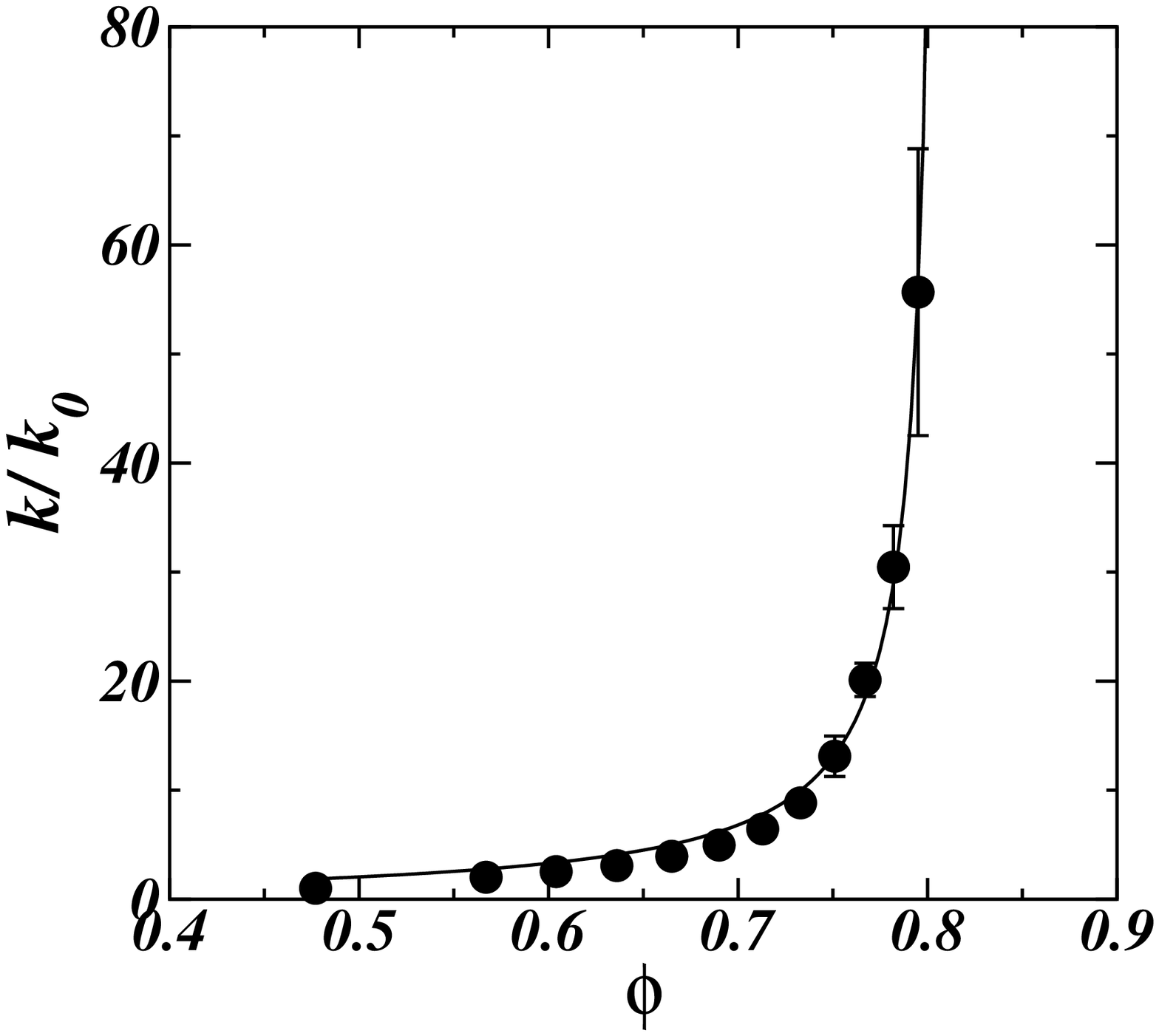}
\caption{Dependence of the macroscopic permeability $\kappa$
at the steady-state on the porosity $\phi$ of the erosion-deposition
system. The solid line is the best fit to the data of the function
$\kappa/\kappa_{0}=(a+b\ln{\phi})^{-1}$, with parameters $a=-0.21$ and
$b=-1.01$.}
\label{kappa_p}
\end{figure}

Finally, it is important to investigate the flow properties of the
porous system in terms of its macroscopic permeability $\kappa$ as
a function of porosity. In Fig.~\ref{kappa_p} we show that, while the
permeability increases very slowly with porosity for diluted systems
(i.e., for low $\phi$ values), it augments in a rather sharp way at
high values of $\phi$ to finally reach a divergent-like behavior in
the vicinity of a porosity close to $\phi=0.8$. Interestingly, we find
that this behavior can be well described by an inverse-logarithmic
relation in the form $\kappa/\kappa_{0}=(a+b\ln{\phi})^{-1}$,
as also depicted in Fig.~\ref{kappa_p}.

\section{Conclusions}

We developed a numerical model to describe the process of
erosion-deposition caused by laminar flow with the drag force given by
the Stokes law. The results we obtained show the formation of a
typical erosion pattern characterized by chevron alignments very
similar to the experimental ones presented in Ref.~\cite{Adrian:2003}
(a typical pattern is shown in Fig.~\ref{experim}). Through
computational simulations performed with this model, we were able to
find dendritic patterns as well as to reproduce the preferential
alignment of the chevron structures observed in real experiments.

Our results indicate that these patterns depend substantially on
the porosity of the system. Previous studies
\cite{Geng:2001,Reyd:2001,William:1989} have shown that the size and shape
of the particles influence dramatically the propagation of the fluid
and the stress distribution in the system. Besides, it has been shown
experimentally that the pattern geometry must also depend on the flow
properties through the porous medium \cite{Andrade:1995,Andrade:1999},
namely, on whether or not the inertial mechanisms of momentum
transport play an important role on the dynamics of pattern
formation. In the present study we considered only laminar flow. By
changing the exponent of the velocity in the drag law, for example,
one can reproduce aspects of turbulent flow to increase the complexity
in the movement of the particles. This could reveal a variety of new
patterns. How the shape and the size distribution of the grains as
well as the flow characteristics affect the patterns are natural
questions that will be addressed in a future work.

In recent works \cite{Igor:2006,Tamas:2005,Malloggi:2006}, many
patterns were observed in experiments involving avalanches, where one
of the most important parameters is the depth of the substrate.
Although the system studied here is related with erosion-sedimentation
processes, this suggests that a variety of different patterns may be
obtained just when one attempts to simulate them in three
dimensions. A simple approximation to a three-dimensional system would
be to consider that, depending on the flux, a particle would not stop
as it reaches another particle, but could jump over it. With this
possibility, the dynamics of the particles changes, as the velocity
of them now depends also on the height of their centers of mass.

\section{Acknowledgements}

We appreciate helpful interactions with A.~M.~C.~Souza,
A.~A.~P.~Olarte, A. A. Moreira, and S.~McNamara. This work was supported by the
Brazilian agencies CNPq, CAPES and FUNCAP and Deutscher Akademischer
Austauschdienst (DAAD). H.~J.~Herrmann thanks the Max-Planck prize.

\bibliography{paper2}

\end{document}